# Democracy & Distrust in an Era of Artificial Intelligence

*Sonia K. Katyal*

*Our legal system has historically operated under the general view that courts should defer to the legislature. There is one significant exception to this view: cases in which it appears that the political process has failed to recognize the rights or interests of minorities. This basic approach provides much of the foundational justifications for the role of judicial review in protecting minorities from discrimination by the legislature. Today, the rise of AI decision-making poses a similar challenge to democracy's basic framework. As I argue in this essay, the rise of three trends – privatization, prediction, and automation in AI – have combined to pose similar risks to minorities. In this essay, I outline what a theory of judicial review would look like in an era of artificial intelligence, analyzing both the limitations and the possibilities of judicial review of AI. Here, I draw on cases in which AI decision-making has been challenged in courts, to show how concepts of due process and equal protection can be recuperated in a modern AI era, and even integrated into AI, to provide for better oversight and accountability.*

Almost forty years ago, in an elegant essay published in *Dædalus*, J. David Bolter wrote, "artificial intelligence is compelling and controversial, not for its practical achievements, but rather for the metaphor that lies behind the programs: the idea that human beings should be seen as nature's digital computers."[1] "The computer," Bolter continued, "is a mirror of human nature, just as any invention reflects to some extent the intellect and character of its inventor. But it is not a perfect mirror; it affects and perhaps distorts our gaze, magnifying certain human capacities … and diminishing others."[2]

As the author points out, a study of AI, which intrinsically compels us to compare mind and machine, reveals the distortions and inaccuracies within each realm. Metaphor, in these contexts, can be a useful way to parse the limits of comparison between humankind and machines. On this point, Bolter wrote, "we do not have to become religious converts to artificial intelligence in order to appreciate the computer metaphor. … Instead, we can ask in what ways the metaphor is apt and in what ways it may fail."[3] In other words, the study of artificial intelligence forces us to examine deep, compositional questions: What makes a hu-man? What makes a machine? And, most important, what makes something artificial, or intelligent?

To some extent, a similar set of compositional comparisons can be posed toward the relationship between law and democracy. Law is a metaphor of sorts – a set of artificial principles – that help us to move toward an ideal society; but the execution of law intrinsically requires us to compare the artifice of these ideals with the unpredictable reality of humanity and governance, thus revealing the distortions and inaccuracies within each realm. Just as computers function as imperfect mirrors of human nature – magnifying certain human capacities and diminishing others – law, too, is a reflection of these limitations and possibilities. And over time, the law has developed its own form of self-regulation to address these issues, stemming from the risks surrounding human fallibility. Our legal system has developed an architectural design of separate institutions, a system of checks and balances, and a vibrant tradition of judicial review and independence. Taken together, these elements compose part of the design of democracy.

Similar elements, I argue in this essay, must be part of the future of artificial intelligence. That is precisely why a study of AI is necessarily incomplete without addressing the ways in which regulation can play a role in improving AI accountability and governance. The issues surrounding algorithmic accountability demonstrate a deeper, more structural tension within a new generation of disputes regarding law and technology, and the contrast between public and private accountability. At the core of these issues, of course, lies the issue of trust: trust in AI, trust in humanity, and trust in the rule of law and governance. Here, the true potential of AI does not lie in the information we reveal to one another, but rather in the issues it raises about the interaction of technology, public trust, and the rule of law.

The rise of AI in decision-making poses a foundational challenge to democracy's basic framework. To recuperate trust in AI for humanity's sake, it is essential to employ design systems that integrate principles of judicial review as a foundational part of AI-driven architecture. My approach in this essay sketches out three dimensions: descriptive, analytic, and normative. First, I describe the background theory of judicial review to introduce a few themes that are relevant to exploring the intersection between AI and our legal system. Then I argue that a system of judicial review is especially needed in light of the rise of three trends that have fundamentally altered the course of AI decision-making: privatization (the increased role of private contractors in making governmental decisions); prediction (the increased focus on using AI to predict human behaviors, in areas as wide-ranging as criminal justice and marketing); and an increased reliance on automated decision-making. These three trends, I argue, have combined to create a perfect storm of conflict that calls into question the role of courts and regulation altogether, potentially widening the gap of protection for minorities in a world that will become increasingly reliant on AI.

Finally, I turn to the normative possibilities posed by these challenges. How can we ensure that software designers, drawn by traditional approaches to statistical, predictive analytics, are mindful of the importance of avoiding disparate treatment? What protections exist to ensure a potential road map for regulatory intervention? Here, drawing on cases in which AI decision-making has been challenged in the courts, I sketch out some ways due process and equal protection can be recuperated in a modern AI era, and even integrated into AI, to provide for better oversight and accountability.

The concept of judicial review, in the United States, has long drawn its force from a famous footnote – perhaps the most famous footnote ever written – in the 1938 case *U.S. vs. Carolene Products*, which involved a constitutional challenge to an economic regulation. In the opinion, written by Justice Harlan Stone, the Court drew a distinction between economic regulation and other kinds of legislation that might affect the interests of other groups. This distinction, buried in that "footnote four," transformed the law's approach to civil rights, underpinning the guarantee of equal protection under the Fourteenth Amendment for all citizens in the future.

For economic regulations, the opinion explained, courts should adopt a more deferential standard of review, erring on the side of trusting the legislature. However, when it was clear that a piece of legislation targeted "discrete and insular minorities," Justice Stone recommended employing a heightened standard of review and scrutiny over the legislation, demanding greater justification to defend its enaction.[4] "When prejudice against *discrete and insular minorities* may be a special condition," Stone wrote, "which tends seriously to curtail the operation of those political processes ordinarily to be relied upon to protect minorities," the law needs to exercise a more "searching inquiry" to justify its actions.

In the footnote, Justice Stone encapsulated a simple, elegant theory: we need the courts to safeguard minorities from regulations that might disregard or disadvantage their interests. Of course, this is not the only reason for why we need judicial review. The famed *Carolene* footnote later formed the backbone of a seminal book by John Hart Ely, *Democracy and Distrust: A Theory of Judicial Review*. Ely's work was essentially a longer explication of this idea: by integrating a healthy distrust of the political process, we can further safeguard democracy for the future. To say that the work is formative would be an understatement, as *Democracy and Distrust* has been described as "the single most cited work on constitutional law in the last century," and "a rite of passage" for legal scholars.[5] By developing the ideas embodied in Stone's footnote, Ely put forth a theory, known as "representation-reinforcement theory," which posits that courts should generally engage in a variety of situations, including cases in which it appears that the political process has failed to recognize the rights or interests of minorities, or where fundamental rights are at stake. This basic theory provides much of the foundational thinking for justifying the role of the judiciary in protecting minorities from discrimination and charting a course for judicial review.

Ely's work has been interpreted to offer a vision of democracy as a function of procedural values, rather than substantive ones, by focusing on the way that judicial systems can create the conditions for a fair political process.[6] One example of this sort of process malfunction, Ely described, involved an intentional kind of disenfranchisement: "the ins," he observed, "are choking off the channels of political change to ensure that they will stay in and the outs will stay out."[7] A second kind of malfunction involved situations in which "no one is actually denied a voice or a vote," but representatives of a majority still systematically disadvantage minority interests "out of a simple hostility or prejudiced refusal to recognize commonalities of interest, and thereby denying that minority the protection afforded other groups by a representative system."[8]

Judicial review, under this approach, also exhorts us to explore whether particular groups face an undue constraint on their opportunity to participate in the political process.[9] For example, if minorities (or other groups) are constrained from participating fully in the political process, then the theory of representation-reinforcement focuses on proxy participation as a solution. Here, Ely reasoned, judges might stand in the place of minorities to ascertain the impact that they may face and take on the responsibility to craft a more inclusive solution. Or if fundamental rights are under threat, the Court should also intervene in order to preserve the integrity of the political process.

This basic theory undergirds much of the institutional and legal relationships between constitutional entitlements and the role of judges in this process. Like any other theory, Ely's approach is not perfect: it has been criticized, and rightfully so, for focusing too much on process at the expense of substantive constitutional rights.[10] But this theory of judicial review also yields both descriptive and normative insights into the government regulation of AI.

R eading Stone's and Ely's concerns in today's era of AI, one is immediately struck by their similarity of context. Both were concerned with the risk of majoritarian control, and designed systems of judicial review to actively protect minority interests. Today, those same concerns are almost perfectly replicated by certain AI-driven systems, suggesting that here, too, judicial review may be similarly necessary. And, normatively, just as judicial review is prescribed as a partial solution to address these risks of majoritarian control in a constitutional democracy, this insight holds similar limits and possibilities in the context of AI regulation.

Put another way, just as our political system often fails to represent the interests of demographic minorities, AI systems carry the same risks regarding the ab-sence of representation and participation – but in private industry. Consider, for example, that one of the most central causes of biased outcomes in AI stems from an underlying problem of lack of representation among minority populations in the data sets used to train AI systems. Machine learning algorithms are, essentially, inherently regressive: they are trained on a body of data that is selected by designers or by past human practices. This process is the "learning" element in machine learning; the algorithm learns, for example, how to pair queries and results based on a body of data that produced satisfactory pairs in the past.[11] Thus, the quality of a machine learning algorithm's results often depends on the comprehensiveness and diversity of the data that it digests.[12]

As a result, bias in AI generally surfaces from these data-related issues of representation.[13] One problem, as AI scholars Kate Crawford and Meredith Whittaker have described, is largely internal to the process of data collection: errors in data collection, like inaccurate methodologies, can cause inaccurate depictions of reality.[14] This absence of representation is a profound cause of the risk of bias in AI. A second issue of bias comes from an external source. It happens when the underlying subject matter draws on information that reflects or internalizes some forms of structural discrimination and thus biases the data as a result.[15] Imagine, for example, a situation in which data on job promotions might be used to predict career success, but the data were gathered from an industry that systematically promoted men instead of women.[16] While the first kind of bias can often be mitigated by "cleaning the data" or improving the methodology, the latter might require interventions that raise complex political ramifications because of the structural nature of the remedy that is required.[17]

As a result, bias can surface in the context of input bias (when the source data are biased because they may lack certain types of information), training bias (when bias appears in the categorization of the baseline data), or through programming bias (when bias results from an AI system learning and modifying itself from incorporating new data).[18] In addition, algorithms themselves can also be biased: the choices that are made by humans – what features should be used to construct a particular model, for example – can comprise sources of inaccuracy as well.[19] An additional source of error can come from the training of the algorithm itself, which requires programmers to decide how to weigh sources of potential error.[20]

All the prior harms may seem representational in nature, but they cause discriminatory effects. If the prior discussion focused on the risks of exclusion from statistical and historical underrepresentation in a data set, there is also the opposite risk of overrepresentation, which can lead to imprecise perceptions and troubling stereotypes. In these instances, due in part to overrepresentation in the data set, an algorithmic model might associate certain traits with another unrelated trait, triggering extra scrutiny. In such cases, it can be hard to prove discrimina-tory intent in the analysis; just because an algorithm produces a disparate impact on a minority group, it does not always mean that the designer intended this result.[21]

Even aside from concerns about data quality and representation, a second cluster of issues emerges from the intersection of privatization and AI-driven governance. Constitutional law scholar Gillian Metzger has presciently observed that "privatization is now virtually a national obsession."[22] Her work describes a foundational risk that private industry is taking the lead in designing modes of governance.[23] Notably, private contractors exercise a broad level of authority over their program participants, even when government officials continue to make determinations of basic eligibility and other major decisions.[24] These trends toward privatization and delegation are endemic throughout government infrastructure, and many draw on machine learning techniques.[25] As intellectual property law scholar Robert Brauneis and information policy law scholar Ellen Goodman have eloquently noted, "the risk is that the opacity of the algorithm enables corporate capture of public power."[26]

Today, algorithms are pervasive throughout public law, employed in predictive policing analysis, family court delinquency proceedings, tax audits, parole decisions, DNA and forensic science techniques, and matters involving Medicaid, other government benefits, child support, airline travel, voter registration, and educator evaluations.[27] The Social Security Administration uses algorithms to aid its agents in evaluating benefits claims; the Internal Revenue Service uses them to select taxpayers for audit; the Food and Drug Administration uses algorithms to study patterns of foodborne illness; the Securities and Exchange Commission uses them to detect trading misconduct; local police departments employ their insights to predict the emergence of crime hotspots; courts use them to sentence defendants; and parole boards use them to decide who is least likely to reoffend.[28]

As legal scholar Aziz Huq has explained, the state uses AI techniques for targeting purposes (that is, decisions on who to investigate or how to allocate resources like aid) and for adjudicatory purposes (in which the state may rely on AI techniques as a stand-in for a judicial determination).[29] To these two parameters, we might add on a third, involving AI-driven forensic techniques to aid the state in determining whether a legal violation has taken place: for example, machine learning techniques that analyze breath alcohol levels. In such cases, while AI might aid the state in gathering evidence, the ultimate determination of compliance (or lack thereof) may rest with human judgment. Here, the selection of a perpetrator might be performed by human law enforcement (who also determine whether evidence supports that a violation has taken place), but the evidence might be informed by an AI-driven technique.

Many of these tools are privately developed and proprietary. Yet the rise of proprietary AI raises a cluster of issues surrounding the risk of discrimination:

one involving the deployment of AI techniques by private entities that raises legal concerns; and another involving the deployment of AI techniques by public entities that raises constitutional concerns. Taken together, these systems can often impose disparate impacts on minority communities, stemming from both private and public reliance on AI. In one example from Pennsylvania, an automated system called the Allegheny Family Screening Tool was used to determine which families were in need of child welfare assistance. But the system entailed the risk of racial disparity: since Black families were more likely to face a disproportionately higher level of referrals based on seemingly innocuous events (like missing a doctor appointment), they were likely to be overrepresented in the data. Parents also reported feeling dehumanized within the system by having their family history reduced to a numerical score. Moreover, given the large amount of data the system processed (and the sensitivity of the data), it carried a serious risk of data breaches.[30]

Each of these prior concerns, as Huq points out, maps onto concerns regarding equality, due process, and privacy, and yet, as he notes, each problem is only "weakly constrained by constitutional norms."[31] Not only would it be difficult to determine whether someone's rights were violated, but parties who were singled out would find it difficult to claim violations of equality, due process, or privacy, especially given the deference enjoyed by the decision-maker.[32] Further, the opacity of these systems raises the risk of (what I have called elsewhere) "information insulation," which involves an assertion of trade secret protection in similar cases.[33]

Each layer of AI-driven techniques raises profound questions about the rule of law. Here, privatization and automation become intimately linked, often at the cost of fundamental protections, like due process. The problem is not just that governmental decision-making has been delegated to private entities that design code; it is also the reverse situation, in which private entities have significant power that is not regulated by the government. While the effects of algorithms' predictions can be troubling in themselves, they become even more problematic when the government uses them to distribute resources or mete out punishment.[34] In one representative case, a twenty-seven-year-old woman with severe developmental disabilities in West Virginia had her Medicaid funds slashed from $130,000 to $72,000 when the vendor began using a proprietary algorithm, making it impossible for her to stay in her family home.[35] When she challenged the determination on grounds of due process, the court agreed with her position, observing that the vendor had failed to employ "ascertainable standards," because it provided "no information as to what factors are incorporated into the APS algorithm," nor provided an "individualized rationale" for its outcome.[36] The district court concluded that the lack of transparency created an "unacceptable risk of arbitrary and 'erroneous deprivation[s]' of due process."[37]

As the previous example suggests, while automation lowers the cost of decision-making, it also raises significant due process concerns, involving a lack of notice and the opportunity to challenge the decision.[38] Even if the decisions could be challenged, the opacity of AI makes it nearly impossible to discern all of the variables that produced the decision. Yet our existing statutory and constitutional schemes are poorly crafted to address issues of private, algorithmic discrimination. Descriptively, AI carries similar risks of majoritarian control and systemic prejudice, enabling majority control at the risk of harming a minority. And yet our existing frameworks for regulating privacy and due process cannot account for the sheer complexity and numerosity of cases of algorithmic discrimination. In part because of these reasons, private companies are often able to evade statutory and constitutional obligations that the government is required to follow. Thus, because of the dominance of private industry, and the concomitant paucity of information privacy and due process protections, individuals can be governed by biased decisions and never realize it, or they may be foreclosed from discovering bias altogether due to the lack of transparency.

I f we consider how these biases might surface in AI-driven decision-making, we can see more clearly how the issue of potential bias in AI resembles the very problem of majority control that Ely wrote extensively about, even though it involves privatized, closed, automated decision-making. If our systems of AI are driven by developers or trained on unrepresentative data, it feeds into the very risk of majoritarian control that judicial review is ideally designed to prevent. I want to propose, however, another story, one that offers us a different set of possibilities regarding the building of trust by looking, again, to the prospect of judicial review.[39] Here, I want to suggest that AI governance needs its own theory of representation-reinforcement, extending to every person within its jurisdiction the equal protection of the law, in essentially the same way that the Constitution purports to.

Where metrics reflect an inequality of opportunity, we might consider employing a similar form of external judicial review to recommend against adoption or refinement of these metrics. In doing so, an additional layer of judicial or quasi-judicial review can serve as a bulwark against inequality, balancing both substantive and process-oriented values. Here, we might use judicial review, not as a tool to honor the status quo, but as a tool to demand a deeper, more substantive equality by requiring the employment of metrics to address preexisting structural inequalities. And if filing an actual legal case in the courts proves too difficult due to an existing dearth of regulation, then I would propose the institution of independent, quasi-judicial bodies to ensure oversight for similar purposes.

What would a representation-reinforcement theory – or relatedly, a theory of judicial review – accomplish in the context of AI? While a detailed account of representation and reinforcement is hard to accomplish in a short essay, I want to focus on two main sets of possibilities, the first stemming from Ely's concept of virtual representation. As I suggested earlier, one core issue with algorithmic decision-making is that it reflects an inherently regressive presumption: decisions, and data collected by past practices, adequately reflect – and predict – what we should do in the future, thereby "freezing" the possibility of a deeper and more meaningful form of substantive equality.[40] Unrepresentative data, in other words, can perpetuate inequalities through machine learning, leading to a feedback loop that further amplifies existing forms of bias.

Interestingly, Justice Stone and John Hart Ely identified roughly the same concerns regarding the lack of minority representation in the democratic pool, justifying a more aggressive form of intervention and oversight. In other words, just as Ely's theory predicts, disparities in representation – over- or underrepresentation – can fuel disparate results. Yet Ely's raising of the "judicial enforceable duty of virtual representation" enables us to see how profitably it can be recast to enfranchise the interests of minority populations in an AI-driven context. As Ely observed, one basic concern is that minorities must always be represented in the political process, and that we rely essentially on our judicial system to make sure that this happens.[41]

Here, one core element to accomplish this goal involves the necessity of creating a layer of institutional separation between the initial decision-maker (the AI system) and the reviewer (essentially, the system of judicial review). Like the division between the judiciary and the legislative branches, AI-driven systems can and must include systems of independent oversight that are distinct from the AI systems themselves. And there is evidence that this architectural solution is taking place. Consider an analogy from Europe's General Data Protection Regulation (GDPR), which requires separate data protection impact assessments (DPIA) whenever data processing "is likely to result in a high risk to the rights and freedoms of natural persons."[42] Large-scale data processing, automated decision-making, processing of data concerning vulnerable subjects, or processing that might prevent individuals from exercising a right or using a service or contract would trigger a DPIA requirement.[43] Notably, this model extends to both public and private organizations.[44]

One could easily imagine how this concept of independent review could be incorporated more widely into AI-driven systems to ascertain whether a system risks disparate impacts. A close look at these statements reveals a markedly thorough implementation of the concept of institutional separation: a DPIA statement is meant to be drafted by the organization's controller in order to show compliance with the GDPR; but the controller represents a separate entity from the organization processing the data.[45] In doing so, the system ensures a form of built-in virtual representation and review by putting the controller in the same position as a judge to ensure compliance. Additional elements require an assessment of risks to individuals and a showing of the additional measures taken to mitigate those risks.[46]

Lastly, at present, as Ely suggests, judicial review is often necessary to ensure due process. Due process is especially needed in the context of AI so that individuals are able to ascertain the rationale behind AI-driven decisions and to guard against unclear explanations. In one case, in Houston, a group of teachers successfully challenged a proprietary algorithm developed by a private company, SAS, called the Educational Value-Added Assessment System (EVAAS) to assess public school teacher performance, resulting in the dismissal of twelve teachers with little explanation or context.[47] Experts who had access to the source code concluded that the teachers were unable to "meaningfully verify" their scores under EVAAS.[48] Ultimately, the court ruled against adopting use of the software because of due process concerns, noting, tellingly: "When a public agency adopts a policy of making high stakes employment decisions based on secret algorithms incompatible with minimum due process, the proper remedy is to overturn the policy."[49] Plainly, the court agreed with the due process concerns, noting that the generalized explanation was insufficient for an individual to meaningfully challenge the system's determination, and the case settled a few months later.[50]

The Houston case is instructive in underscoring the importance of safeguarding procedural protections like due process. Had it not been for the teachers' ability to bring this to a judicial forum to demand due process protection, the AI-driven injustice they faced would have never seen the light of day. By requiring AI systems to integrate similar entitlements of due process and independent oversight, we can ensure better outcomes and build more trust into the accountability of AI-driven systems overall.

In his essay forty years ago, Bolter predicted, "I think artificial intelligence will grow in importance as a way of looking at the human mind, regardless of the success of the programs themselves in imitating various aspects of human thought. . . . Eventually, however, the computer metaphor, like the computer itself, will simply be absorbed into our culture, and the artificial intelligence project will lose its messianic quality."[51]

We are still at a crossroads in adapting to AI's messianic potential. Ely wrote his masterful work at a time in which AI was just at the horizon of possibility. Yet the way that AI promises to govern our everyday lives mirrors the very same concerns that he was writing about regarding democracy and distrust. But the debates over AI provide us with the opportunity to elucidate how to employ AI to build a better, fairer, more transparent, and more accountable society. Rather than AI serving as an obstacle to those goals, a robust employment of the concept of judicial review can make them even more attainable.

AUTHOR'S NOTE

The author thanks Erwin Chemerinsky, James Manyika, and Neal Katyal for their insightful comments and suggestions.

ABOUT THE AUTHOR

**Sonia K. Katyal** is Associate Dean of Faculty Development and Research, Codirector of the Berkeley Center for Law & Technology, and Distinguished Haas Chair at the University of California, Berkeley, School of Law.

ENDNOTES

[1] J. David Bolter, "Artificial Intelligence," *Dædalus* 113 (3) (Summer 1984): 3.

[2] Ibid., 17.

[3] Ibid.

[4] *U.S. vs. Carolene Products Co.*, 304 U.S. 144, 153 n.4 (1938).

[5] Henry Paul Monaghan, "John Ely: The Harvard Years," *Harvard Law Review* 117 (6) (2004): 1749.

[6] Jane S. Schacter, "Ely and the Idea of Democracy," *Stanford Law Review* 57 (3) (2004): 740.

[7] John Hart Ely, *Democracy and Distrust: A Theory of Judicial Review* (Cambridge, Mass.: Harvard University Press, 1980), 103.

[8] Schacter, "Ely and the Idea of Democracy," 740.

[9] See Ely, *Democracy and Distrust*, 77, quoted in Schacter, "Ely and the Idea of Democracy," 741.

[10] See, for example, Lawrence Tribe, "The Puzzling Persistence of Process Based Theories," *Yale Law Journal* 89 (1063) (1980).

[11] See Schacter, "Ely and the Idea of Democracy," 760.

[12] See Andrew D. Selbst and Solon Barocas, "Big Data's Disparate Impact," *California Law Review* 104 (3) (2016): 688.

[13] Kate Crawford and Meredith Whittaker, "The AI Now Report: The Social and Economic Implications of Artificial Intelligence Technologies in the Near-Term" (New York: AI Now Institute, July 7, 2016, last modified September 22, 2016), 6–7, https://ainowinstitute.org/AI_Now_2016_Report.pdf [http://perma.cc/6FYB-H6PK (captured August 13, 2018)].

[14] Ibid.

[15] Ibid.

[16] Ibid., 6.

[17] Ibid.

[18] Nizan Geslevich Packin and Yafit Lev-Aretz, "Learning Algorithms and Discrimination," in *Research Handbook on the Law of Artificial Intelligence*, ed. Ugo Pagallo and Woodrow Barfield (Cheltenham, United Kingdom: Edward Elgar Publishing, 2018), 88–133.